\documentclass[]{aa}  

\usepackage{graphicx}
\usepackage{txfonts}
\usepackage{hyperref}

\usepackage{amsmath}
\usepackage{upgreek}
\usepackage[nopatch]{microtype}
\usepackage{booktabs}
\usepackage{multirow}
\usepackage{arydshln}
\usepackage{siunitx}
\usepackage{placeins}

\newcommand{\g}{G321.3--3.9} 

\newcommand{\fig}{Fig.}

\newcommand{\sect}{Section}
\newcommand{\sects}{Sections}

\newcommand{\tab}{Table}

\begin{document} 

\title{G321.3--3.9: A new supernova remnant observed with multi-band radio data and in the SRG/eROSITA All-Sky Surveys}

\author{S. Mantovanini\inst{1}
    \and
    W. Becker\inst{2}\inst{,3}
    \and
    A. Khokhriakova\inst{2}
    \and
    N. Hurley-Walker\inst{1}
    \and
    G. E. Anderson\inst{1}
    \and
    L. Nicastro\inst{4}
}

\institute{International Centre for Radio Astronomy Research, Curtin University, Bentley WA 6102, Australia\\
    \email{silvia.mantovanini@postgrad.curtin.edu.au}
    \and
    Max-Planck-Institut f\"{u}r extraterrestrische Physik, Gie{\ss}enbachstra{\ss}e 1, D-85748 Garching, Germany
    \and
    Max-Planck-Institut für Radioastronomie, Auf dem Hügel 69, 53121 Bonn, Germany 
    \and
    INAF - Osservatorio di Astrofisica e Scienza dello Spazio di Bologna, via Piero Gobetti 93/3, I-40129 Bologna, Italy
}

\date{Received ; accepted }
 
\abstract
{}
{\g\ was first identified as a partial shell at radio frequencies a few decades ago. Although it continued to be observed, no additional studies were undertaken until recently. 
}
{
In this paper we present results from a large selection of radio and X-ray data that cover the position of \g. We confirmed \g\ as a new supernova remnant (SNR) using data collected by several radio surveys, spanning a frequency range from 200 to 2300\,MHz. Stacked eROSITA data from four consecutive all-sky surveys (eRASS:4) provide spectro-imaging information in the energy band $0.2-8.0$ keV.
}
{
\g\ has an elliptical shape with major and minor axes of approximately $1\fdg7 \times 1\fdg1$. From CHIPASS and S-PASS data, we calculate a spectral index $\alpha = -0.8 \pm 0.2$, consistent with synchrotron emission from an expanding shell in the radiative phase. The eROSITA data show an X-ray diffuse structure filling almost the entire radio shell. 
Based on our spectral analysis, we found the temperature to be approximately 0.6 keV and the column absorption density about $10^{21}$ cm$^{-2}$. Comparing this absorption density to optical extinction maps, we estimated the distance to fall within the range of (1.0 -- 1.7) kpc, considering the $1 \sigma$ uncertainty range.
}
{}

\keywords{ISM: individual objects: \g, ISM: supernova remnants, radiation mechanisms: non-thermal}

\titlerunning{Confirmation of G321.3-3.9 as a supernova remnant}
\authorrunning{Mantovanini et al.}
 
\maketitle

\section{Introduction} \label{sec:intro}
Supernova remnants (SNRs) \citep[see][for a detailed review]{Dubner2015} significantly contribute to the chemical and physical evolution of the Galaxy. The material and energy released during an explosion interact with the interstellar medium (ISM), generating a shock wave that expands more or less spherically depending on the ISM density and the evolutionary phase of the remnant \citep[originally described by ][]{Woltjer1972}. 

A census of SNRs is regularly updated by \citet{Green2019}.\footnote{https://www.mrao.cam.ac.uk/surveys/snrs/} Confirmed and candidate sources are included along with a summary of their main properties:  position, flux density ($S\propto \nu^\alpha$), size, and spectral index. The December 2019 version contains 294 confirmed objects and $\simeq$~300 possible candidates detected in radio, optical, X-ray, and $\gamma$-ray. Most of the sources we know have been identified using radio observations where electrons lose energy slowly and remain visible for up to 150000 years. Recently, modern instruments such as the Murchison Widefield Array \citep[MWA;][]{Tingay2013,Wayth2018}, the Australian Square Kilometre Array Pathfinder (ASKAP), and the Low-Frequency Array (LOFAR) have improved the sensitivity and resolution of
 surveys and have enabled the identification of new low surface brightness SNRs such as Hoinga \citep{Becker2021}, G118.4+37.0 \citep{Arias2022}, and G288.8–6.3 \citep{Filipovic2023}. The objects are all located at high Galactic latitudes with a confirmed counterpart at X-ray and/or $\gamma$-ray energies, which provided us with a clearer understanding of the ongoing processes within the source.

In this paper we illustrate the identification of   SNR \g, which appears as an extended structure filling the radio shell,\ via multi-band radio data and its simultaneous detection in X-ray data. \g\ has been detected and classified as a SNR candidate three times. The first time was by \cite{Duncan1997} using a polarimetric radio continuum survey at 2.4\,GHz taken with the Parkes 64m radio telescope. The survey covers a region of the sky between $\ang{238} < l < \ang{365}$ and $|b| < \ang{5}$ with a resolution of 10.4$'$. The authors classified the object as a partial elliptical structure with a size of $1\fdg8 \times 1\fdg3$, showing a well-defined arc on the southeastern side. 
\g\ was detected for the second time during the second epoch Molonglo Galactic Plane Survey  \citep[MGPS-2; ][]{Green2014} at 843\,MHz, as one of the 23~SNR candidates the authors identified along the Galactic plane. They described \g\ as an elliptical and almost complete shell with size of $109 \times 64\, \text{arcmin}^{2}$ ($1\fdg8 \times 1\fdg1$), peak flux of $10 \, \text{mJy/arcmin}^{2}$, and total integrated flux density greater than $0.37 \; \text{Jy}$. For the third detection, a deep optical RGB composed of \textsc{[O iii]}, H$_{\alpha}$, and \textsc{[S ii]} emission-line images of \g\ was reported by \citet{Fesen2024}, in which the remnant structure reveals strong optical filaments. The detected emission lines typically characterize SNR regions where radiative shocks occur, in other words, shocks that have significantly reduced their velocity ($<200$\,km/s) and are consequently associated with old objects. Here we confirm the identification of this SNR through the use of modern surveys conducted with the MWA at 200\,MHz and the eROSITA instrument \citep{Predehl2021} on board the Russian--German Spectrum-Roentgen-Gamma spacecraft \citep{Sunyaev2021}, as well as other archival and publicly accessible radio surveys. 

The structure of the paper is as follows. The radio observations and analysis are described in \sect~\ref{sec:radio}, including archival data for calculating the spectral index of the SNR. In \sect~\ref{sec:x-rays} we describe the eROSITA X-ray detection and analysis. In \sects~\ref{sec:summary} and ~\ref{sec:conclusions} we summarize the results and main properties of SNR \g\ and present our  conclusions.

\section{Radio data} \label{sec:radio}
Following the identification of the candidate in the Galactic Plane Monitoring (GPM) data (see \sect~\ref{sub:mwa}), we searched all other archival surveys to detect this SNR candidate. In addition to MGPS-2, we identified it in the Continuum HI Parkes All-Sky Survey \citep[CHIPASS; ][]{Calabretta2014}, and the S-band Polarisation All Sky Survey \citep[S-PASS; ][]{Carretti2019}, which we used in the analysis for this paper and are described below. It was also detected in the Rapid ASKAP Continuum Survey \citep[RACS;][]{Mcconnell2020} and the Evolutionary Map of the Universe \citep[EMU][]{Norris2021}, although the data was unsuitable for our analysis. For completeness, the images from these two surveys can be found respectively in Appendix \ref{racs} and \ref{emu}.

\subsection{GPM} \label{sub:mwa}
From July to September 2022, a  GPM campaign across $185 - 215$\,MHz (see Hurley-Walker et al., in prep., for additional information) was conducted with the MWA, covering approximately one-third of the southern Galactic plane with a bi-weekly cadence. The data collected were calibrated, imaged, and mosaicked, achieving a sensitivity of $\simeq 1-2$\;mJy/beam, extremely low compared to previous radio surveys of the Galactic plane at MHz frequencies, as well as an angular resolution of $45''$. The aim of the campaign was to identify transient sources in the Milky Way, particularly the long-period radio transients \citep{2023Natur.619..487H}. However, the information can be transferred to other science goals, such as studying SNRs. The observed sky area covers $\ang{285} < l < \ang{70}$ and $|b| < \ang{16}$ for a total of $4600$\,degrees squared.

Using this image, it was possible to identify 21~SNR candidates, one of which is \g. The sampling criteria correspond to a first selection based on the morphology of the objects, typically a ring-shaped structure that could be more or less elongated depending on the condition of the medium surrounding it. Second, we compared this with infrared emission, which is essential for distinguishing between thermal and nonthermal sources. Commonly, thermal regions (particularly \textsc{Hii} regions) can be incorrectly classified as SNRs due to the similarity in their morphology at radio frequencies. However, \textsc{Hii} regions show a distinctive structure in the infrared bands, where an annulus region of polycyclic aromatic hydrocarbon radiation (dominant at $\simeq 10 \, \text{$\mu$m}$) surrounds the hot dust radiation at $24 \, \text{$\mu$m}$. In this paper we present this particular SNR's radio and X-ray properties.

\begin{figure*}
\centering
\includegraphics[width=1.0\linewidth]{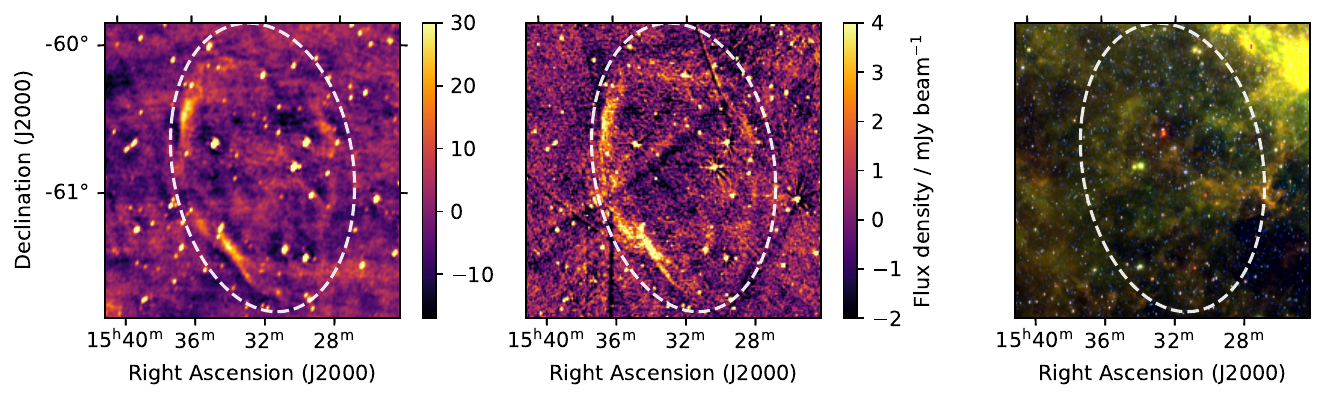}
\caption{$3\fdg5 \times 3\fdg5$ region surrounding \g\ as seen by GPM at 200\,MHz in the left panel and by MGPS-2 at 843\,MHz in the middle panel. An infrared zoomed-in visualization by WISE at 22 $\mu$m (R), 12 $\mu$m (G), and 3.4 $\mu$m (B) is shown in the right panel. The white dashed line highlights the location of  the candidate.}
\label{fig:GPM-MGPS}
\end{figure*}

\g\ appears in the GPM image as an elliptical-shaped radio shell with major and minor diameters equal to $1\fdg7$ and $1\fdg1$, respectively, and an inclination angle of about $\ang{125}$ from the west, as shown in the left panel of Figure~\ref{fig:GPM-MGPS}. 
Due to the lack of short baselines in the survey, it is not possible to completely measure the flux density on large scales for these observations. As can be seen in the left panel of \fig~\ref{fig:GPM-MGPS}, only the left side of the shell appears to be above the background level. Therefore, it is not possible to accurately measure the radio flux density for this particular remnant with GPM data. 

\subsection{MGPS-2}\label{mgps2}
The middle panel of \fig~\ref{fig:GPM-MGPS} shows the detection of \g\ in the MGPS-2, a deep radio survey of the entire sky south of declination $-30$ degrees, made using the \textit{Molonglo Observatory Synthesis Telescope} \citep[MOST][]{Mills1981} at $843 \, \text{MHz}$. MGPS-2 has a resolution that varies with declination: $45^{\text{"}} \times 45^{\text{"}} \csc \, |\delta|$ and a noise level of $\sim1 \, \text{mJy beam}^{-1}$. We downloaded\footnote{http://www.astrop.physics.usyd.edu.au/mosaics/Galactic/} the extended source map to create a $3\fdg5 \times 3\fdg5$ cutout around the SNR candidate. Similarly to the GPM image, the remnant appears as a nonfilled shell, with the left portion of the shell being brighter than the rest of the structure. The MOST telescope does not measure the correlations on baselines shorter than 42.9$\lambda$, making extended sources with an angular scale larger than $\simeq 20 '$ not detectable by the instrument, and therefore it is impossible to calculate the flux density accurately. Only a lower limit can be provided (as mentioned in \sect~\ref{sec:intro}).

\subsection{CHIPASS}\label{subsub:chipass}
CHIPASS covers the entire sky south of declination $+25^{\circ}$ and operates at a central frequency of $1394.5 \, \text{MHz}$ with a sensitivity of $40 \, \text{mK}$. This survey provides an excellent tool for measuring the total power of extended objects. We downloaded\footnote{https://www.atnf.csiro.au/people/mcalabre/CHIPASS/index.html} the extended source map to create a $3\fdg5 \times 3\fdg5$ cutout around the SNR candidate, which we then regridded to match the GPM mosaic (left panel of \fig~\ref{fig:CHIPASS_SPASS}). 

\subsection{S-PASS} \label{subsub:spass}
S-PASS operates at a central frequency of $2303 \, \text{MHz}$ with a sensitivity of $9 \, \text{mK}$. S-PASS covers the entire southern sky with declination lower than $-1^{\circ}$. This survey is particularly useful because it can detect polarized radio emission, given information regarding the magnetic field in the region of the sky under examination. We downloaded\footnote{https://sites.google.com/inaf.it/spass} the Stokes I source map to create a $3\fdg5 \times 3\fdg5$ cutout around the SNR candidate, which we then regridded to match the GPM mosaic (right panel of \fig~\ref{fig:CHIPASS_SPASS}). Stokes Q and U were not used in the principal analysis, but are reported in Appendix \ref{fig:Polarization} for completeness.

\begin{figure*}
\centering
\includegraphics[width=1.0\linewidth]{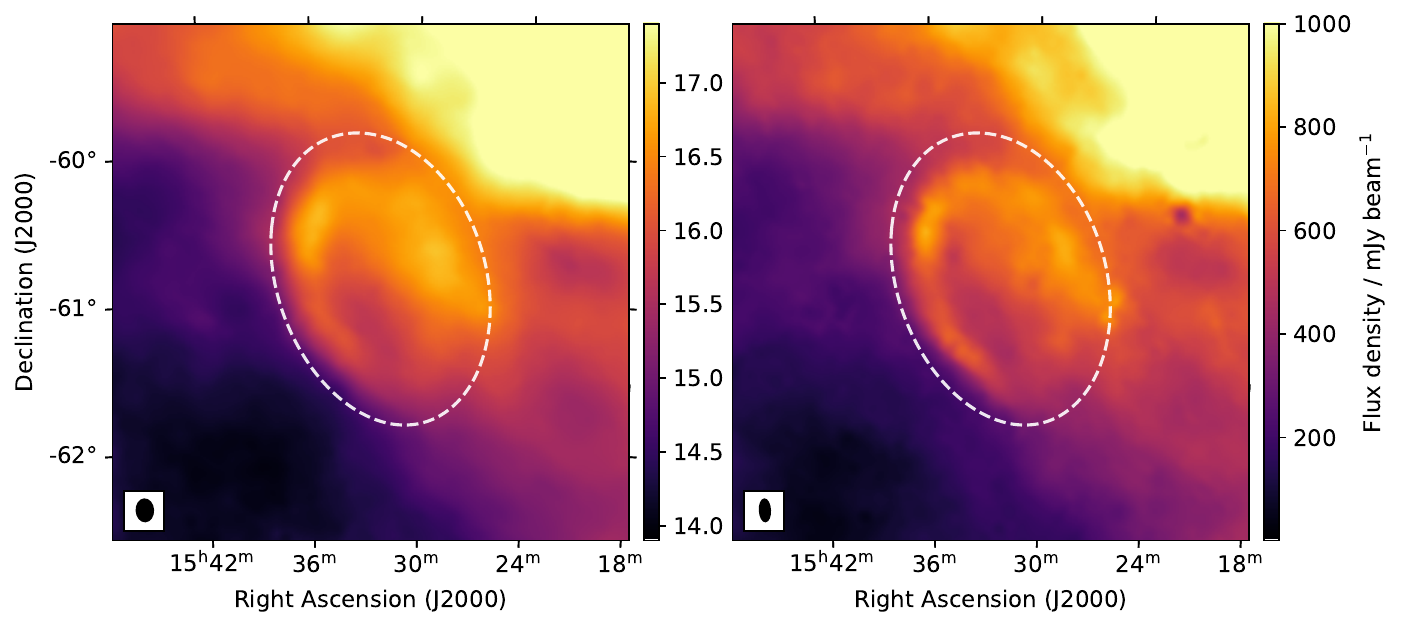}
\caption{$3\fdg5 \times 3\fdg5$ of the region surrounding \g\ as seen at 1.4\,GHz by CHIPASS (left panel) and at 2.3\,GHz by S-PASS (right panel). The panels show the region after the conversion from K to Jy beam$^{-1}$ and the subtraction of point sources. The dashed line highlights the area within which the SNR is located.}
\label{fig:CHIPASS_SPASS}
\end{figure*}

\subsection{Radio spectral analysis} \label{subsec:index}
We now describe how we calculated the spectral index of \g\ using the CHIPASS and S-PASS detections shown in \fig~\ref{fig:CHIPASS_SPASS}. These maps were chosen for this calculation because the single-dish data measure the total flux density of the source and do not ``resolve out'' the structure, as is the case for interferometric measurements. A spectral index of $-1.1 < \alpha < -0.1$ characterizes radio SNR shells \citep{Dubner2015}; thus, a measurement falling in this range will further support the nonthermal nature of \g. 

Initially, we needed to remove any flux density contributions from point sources within the extent of the SNR. This was done by subtracting the point sources at locations identified in the higher-resolution MGPS-2 data from the CHIPASS and S-PASS images. The point sources were identified within $2\fdg3$ of \g\ using the algorithm Aegean, which utilizes background and noise models previously generated with \textsc{BANE} \citep[see][for technical details]{Hancock2012,Hancock2018} to detect significant pixels in the image and to create a catalog of sources giving, among other things, coordinates, peak flux, and integrated flux.

We then applied the three following steps:

\begin{itemize}
\item convolution of the catalog created to match the resolution of CHIPASS and S-PASS; 
\item creation of a FITS image in the same sky frame with the model of the sources to subtract assuming a spectral index value corresponding to the median local value of $-0.83$ \citep[according to the source catalog of the southern sky presented in ][]{Mauch2003} fixing the error to the width of the spectral distribution;
\item subtraction of the model from the original CHIPASS and S-PASS images. 
\end{itemize}

At this point, we converted the CHIPASS and S-PASS image units from Kelvin to Jansky, and we applied the \textsc{polygon\_flux} \citep[][]{Hurley2019a}\footnote{https://github.com/nhurleywalker/polygon-flux} software to measure the flux density of the remnant in both surveys. The polygon selection around the source is the dominant source of error,\footnote{We applied a mean background and an interpolated 2D plane. Since the variation between the two options is 3\%, which is lower than the error caused by the polygon selection effects, we report here only the results obtained using a mean background.} so we measured the flux density of the SNR ten times. The flux density has been assumed to be the average of these measurements, and the uncertainty is calculated as the standard deviation of this data set. The results are reported in \tab~\ref{tab:fluxdendity}; as expected for a nonthermal source, the SNR is brighter at lower frequencies.  

\begin{table} \footnotesize
\caption{Integrated flux densities of \g\ from the CHIPASS and S-PASS detections used to calculate the SNR spectral index.
}
\label{tab:fluxdendity}
\begin{center}
\begin{tabular}{cccc} \toprule
Survey & Frequency & Resolution & Flux density \\ & MHz & arcmin & Jy \\
\midrule
CHIPASS & 1400 & 14.4 & 11.5 $\pm$ 0.9\\ 
S-PASS & 2300 & 8.9 & 7.6 $\pm$ 0.6\\      
\bottomrule \\
\end{tabular}
\end{center}

Notes: \textsc{polygon\_flux} on images where contaminating sources had been removed.

\end{table}

The two data points were fitted with a power law using the propagation of the errors to determine the uncertainty, yielding a spectral index value of $\alpha = -0.8 \pm 0.2$.
We also cross-checked this result with infrared data. The two higher bands (at $22 \, \text{$\mu$m}$ and $12 \, \text{$\mu$m}$) of the all-sky survey release of the \textit{Wide-field Infrared Survey Explorer} \citep[WISE][]{Wright2010} provide images in the sky area of \g. These do not show the typical \textsc{Hii} region morphology of a central region bright at 22$\mu$m and a ring of 12$\mu$m emission from excited PAHs, as shown in the right panel of \fig~\ref{fig:GPM-MGPS}. We thus confirm that the radio emission is from nonthermal radiation.

\section{eROSITA observations of \g} \label{sec:x-rays}
The X-ray data we report here were taken during the first four eROSITA all-sky surveys eRASS:4 \citep{Predehl2021, Sunyaev2021}. \g\ was mapped in 80 telescope passages in the following date intervals: March 6--9 and September 5--8, 2020; February 23--25 and August 30--September 02, 2021. These observations resulted in an averaged vignetting-corrected exposure time of 
710\,s ($\sim$177\,s per sky survey). eROSITA is an array of seven X-ray telescopes observing the same sky region. Each telescope has a detector module. The detectors of telescope modules (TM) 1, 2, 3, 4, and 6  differ from the detectors of modules TM5 and TM7 by having an on-chip filter; hence, each detector has a slightly different detector response. In addition, detector modules 5 and 7 are partially affected by a light leak, depending on the telescope's orientation with respect to the Sun. We verified that this was not a problem for the observation of \g, so we used all the data from eROSITA's seven telescope modules in the X-ray image reconstruction. These data were processed by the eROSITA Standard Analysis Software (eSASS) pipeline \citep{Brunner2022} with the processing number $\#020$. For the data analysis, we used the standard tasks available in eSASS version 211214 (released on December 21, 2021).\footnote{\url{https://erosita.mpe.mpg.de}} The eSASS pipeline divides X-ray data into 4700 partly overlapping sky tiles of $3\fdg6 \times 3\fdg6$ each. These are numbered using six digits, three for RA and three for Dec, representing the sky tile center position in degrees. \g\ falls into the eRASS sky tile numbered 231150.

\begin{figure}
\centering
\includegraphics[width=\columnwidth]{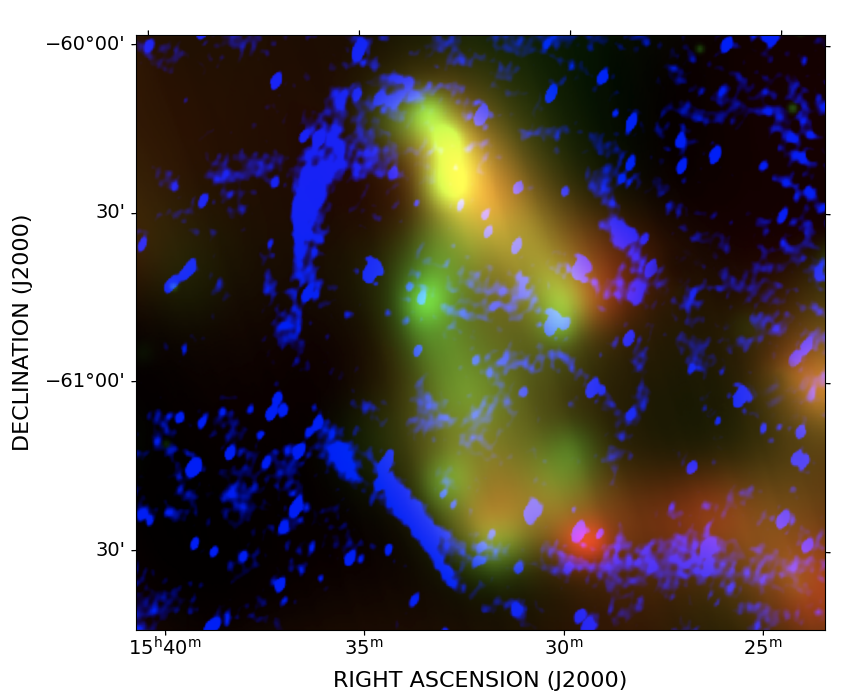}
 \caption{Composite RGB image of \g\ as seen in the eROSITA all-sky surveys eRASS:4 and the MWA radio image taken at 200 MHz. The X-ray photons to produce the image were color-coded according to their energy (red for energies 0.2--0.7, green for 0.7--1.2\;keV), whereas the MWA image of \g\ is shown in the blue channel.}
\label{fig:eROSITA_RGB}
\end{figure}

Figure~\ref{fig:eROSITA_RGB} depicts a composite X-ray--radio RGB image of \g, color-coded according to the energy of the detected photons. To produce the figure, we first created single-band pictures for the three energy bands 0.2--0.7\;keV, 0.7--1.2\;keV, and 1.2--8.0\;keV. 
As there was no significant emission in the 1.2--8.0 \,keV band, we loaded the MWA radio image taken at 200 MHz into the blue channel to create the RGB image shown in Figure ~\ref{fig:eROSITA_RGB}.
eROSITA's field of view had an average angular resolution of $26\arcsec$ during its all-sky survey mode.
The spatial binning in the R- and G-bands of Figure~\ref{fig:eROSITA_RGB} was set to $20\arcsec$ to limit the slight resolution reduction by the smoothing process. 

To enhance the visibility of diffuse X-ray emission to its largest extent, we applied the adaptive kernel smoothing algorithm of \cite{Ebeling2006} with a Gaussian kernel of $3.5~\sigma$ after point-source removal in each of the single-band images. 
We further computed the vignetting corrected exposure maps for the R- and G- eROSITA bands and applied the vignetting correction to the corresponding images. The cut values used for the R- and G-bands are at $(1.53 - 5.47) \times 10^{-5}$ cts/s and $(2.04 - 12.13) \times10^{-5}$ cts/s, respectively. 
As can be seen from Figure~\ref{fig:eROSITA_RGB}, diffuse X-ray emission fills a large part of the inner remnant region, partly overlapping with the remnant's radio emission in the MWA image taken at 200 MHz in the west and southeast. We also found a similar structure in the ROSAT all-sky survey, which has made this source a target in our ROSAT SNR candidate list. As the quality of the eROSITA data supersedes those from the ROSAT all-sky survey, we did not include the ROSAT data in this publication. Galactic absorption, eROSITA's energy response, and the remnant's energy emission spectrum result in a more substantial X-ray flux at medium energies (0.7--1.2\,keV) than in the soft and hard bands. 
In addition we checked the \textit{XMM-Newton} and \textit{Chandra} data archives. We found an 11.6 ks \textit{XMM-Newton} observation covering the northern region of \g. A short report about these data can be found in Appendix \ref{sec.xmm}.


We performed the X-ray spectral analysis using \texttt{PyXSPEC}, the Python implementation of \texttt{XSPEC} \citep{Arnaud1996}. The data extraction was limited to eROSITA TMs 1-4 and 6 because of the potential light leak in     TM5 and TM7, which makes their spectral calibration uncertain. For spectral fitting, we used the Cash statistic \citep{Cash1979} as implemented in XSPEC, presenting errors at $1 \sigma$ confidence levels. The $\mbox{CSTAT}/\mbox{d.o.f.}$ (degrees of freedom) value was employed to assess the quality of the fit. 

To exclude unrelated point source contributions in our spectral analysis to the largest extent possible, we removed all point sources located within the remnant boundaries if they had a logarithmic detection likelihood threshold of DETLIKE $= -\ln({\rm P}) \ge 70$.
This threshold might appear stringent, but it is necessary to avoid the inclusion of numerous spurious point sources, mainly because the source detection algorithm treats diffuse emission as a plethora of individual point sources. The remaining fainter point source contributions detected with a likelihood $< 70$ are considered to be negligible relative to the overall emission of the SNR. All identified point sources were masked out using circular regions with a radius of $120^{\prime\prime}$. The radius was chosen according to the instrument's point spread function.

\begin{figure}
    \centering
    \includegraphics[scale=0.2]{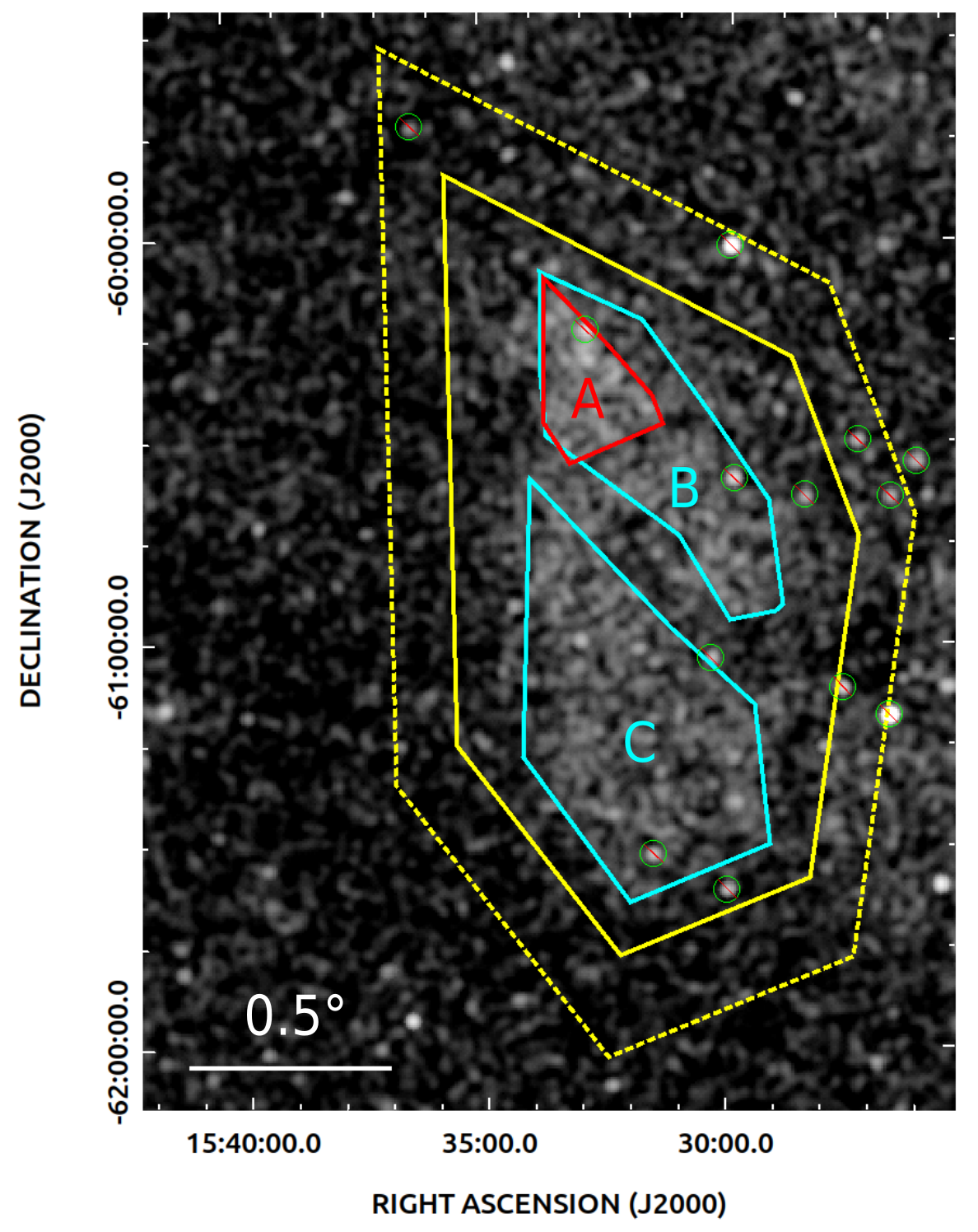}
    \caption{eRASS:4 smoothed image of \g\ in the 0.5 -- 2.4 keV energy range with regions used for the eROSITA X-ray spectral analysis. The main source region is the solid yellow polygon surrounding the SNR. The point sources subtracted from the field are highlighted by a green circle with a red line in the center. The background region is the area between that polygon and the bigger dotted polygon.}
    \label{fig:regions}
\end{figure}

The diffuse X-ray emission of the remnant was extracted from a polygonal region encompassing the remnant with additional analysis in three internal regions labeled A, B, and C (see Fig.~\ref{fig:regions}). The background was extracted from a scaled polygonal region surrounding the source region, approximately $30\%$ larger than the original source-surrounding polygon. This region includes surrounding diffuse X-ray sources to the north and west of the remnant, potentially introducing contamination into the analysis. Therefore, we also tested an alternative background region located to the east of the remnant, where no contaminating X-ray diffuse emission was observed. The results of the spectral analysis remained consistent within the margins of error.

The background spectral model considered the contribution from the instrumental background, a fixed absorbed extragalactic X-ray background, and a thermal background component (see, e.g., \citealt{Mayer2023} for further details). For our source spectral models, the first tested was vpshock, a constant temperature plane-parallel shock plasma model \citep{Borkowski2001}.
The absorption component was modeled with the \texttt{TBabs} model \citep{Wilms2000}. The second model, \texttt{TBabs*vapec}, represents an emission spectrum from collisionally ionized diffuse gas, while the third model, TBabs*vnei, is a nonequilibrium ionization collisional plasma model. 
In all the models, we allowed the abundance of O, Ne, Mg, Si, and Fe to vary, keeping the other elements frozen at solar values.

The limited photon statistics of \g\ in the eROSITA survey data allow only approximate spectral modeling. The models all fitted the data with a comparable goodness of fit, as detailed in Table~\ref{tab:all_models}, reflecting the challenges in distinguishing between different physical emission scenarios. 

Among the tested spectral models, the temperature and the absorption column density ($N_{\mathrm{H}}$)  were found to be consistent with values of approximately 0.6 keV and $10^{21}$ cm$^{-2}$, respectively, within a 2 $\sigma$ error range. 
While temperatures show consistently small errors among all fitted models, abundances are determined with larger uncertainties due to the limited photon statistics. 
A noticeable scarcity of Ne and Mg is common to all fitted spectral models. Unlike Mg, Ne cannot be depleted into dust grains \citep{Hwang2008}. 
To test the spectral results for their dependencies on the abundances of these elements, we fixed Ne and/or Mg to a solar value. Specifically, when fixing both Ne and Mg, CSTAT deteriorated by $ \Delta C = 44$ for \texttt{vpshock}, $ \Delta C = 16$ for \texttt{vapec}, and $ \Delta C = 17$ for \texttt{vnei}; the errors of the fitted abundances of the other elements became larger, the normalization decreased by a factor of 10, and the temperature increased to a high value of 1 keV. Fixing only Ne led to similar effects.

\begin{figure}
    \centering
    \includegraphics[scale=0.3]{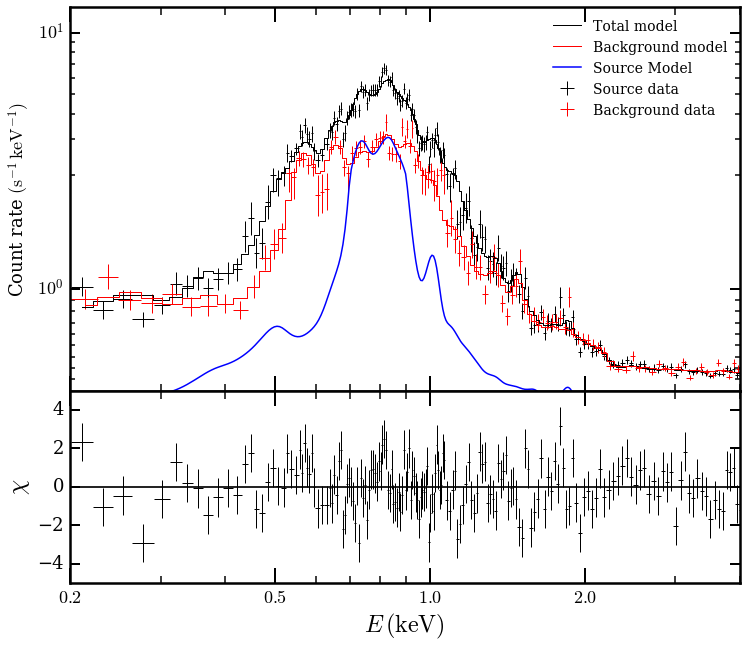}
    \caption{X-ray spectrum of \g\ obtained with eROSITA. The fit model is TBabs*vapec.}
    \label{fig:X_spectrum}
\end{figure}

Assuming that the X-ray emission is dominated by emission from the ejecta, this may suggest a thermonuclear supernova origin, as core-collapse supernovae typically exhibit a higher ratio of Ne and Mg to Fe (see, e.g., Fig. 2.8 in \citealt{Vink2020}). 
However, determining whether the X-ray emission is dominated by the ejecta or by circumstellar and/or interstellar material requires further analysis.
The spectral fit obtained using the absorbed \texttt{vapec} model is shown in  Fig.~\ref{fig:X_spectrum}, whereas the fitting results of all tested models are listed in Table~\ref{tab:all_models}.

The fitting results for the subregions demonstrate notable consistency in temperature and absorption (see Table~\ref{tab:spectra-regions}). 
The temperature remains around 0.6 keV and hydrogen column density $N_{\mathrm{H}} \sim 10^{21}$ cm$^{-2}$ for the \texttt{vapec} model, with variations within the 1 $\sigma$ error range. 

It should be noted that, from the physical point of view, the \texttt{vpshock} model is likely not applicable in the case of \g\ since it is probably an older remnant in the radiative phase, as is discussed below. However, we present the results of all tested spectral models for completeness and reference of the reader.

\begin{table*}[]
\centering
\caption{ Best fit parameters with $1 \sigma$ errors obtained with different models;    O, Ne, Mg, Si, and Fe were left free to vary.
\label{tab:all_models}}
\begin{tabular}{cccc}
\toprule
Model & \texttt{vpshock} & \texttt{vapec} & \texttt{vnei} \\
\midrule
$N_{\mathrm{H}}$ (10$^{22}$ cm$^{-2}$) & $0.15_{-0.03}^{+0.04}$ & $0.11_{-0.02}^{+0.02}$ & $0.12_{-0.02}^{+0.03}$ \\[1ex]
kT (keV) & $0.65_{-0.03}^{+0.03}$ &  $0.58_{-0.02}^{+0.02}$ &  $0.58_{-0.03}^{+0.02}$  \\[1ex]%
O/O$_{\odot}$ & $0.17_{-0.05}^{+0.08}$  & $0.18_{-0.11}^{+0.15}$ & $0.08_{-0.08}^{+0.08}$ \\[1ex]%
Ne/Ne$_{\odot}$ & $< 0.05$ & $0.3_{-0.12}^{+0.18}$ & $0.17_{-0.06}^{+0.08}$  \\[1ex]%
Mg/Mg$_{\odot}$ & $< 0.05$  &  $< 0.05$ &  $0.06_{-0.06}^{+0.09}$ \\[1ex]%
Si/Si$_{\odot}$ & $0.6_{-0.2}^{+0.2}$ & $0.51_{-0.23}^{+0.14}$ & $0.6_{-0.2}^{+0.3}$ \\[1ex]%
Fe/Fe$_{\odot}$ & $0.5_{-0.1}^{+0.2}$ &  $0.30_{-0.05}^{+0.07}$ &  $0.37_{-0.07}^{+0.09}$ \\[1ex]%
$\tau_{u}$ (10$^{10}$ cm$^{-3}$ s) & $51_{-26}^{+75}$ & - &  $17_{-4}^{+11}$ \\[1ex]%
Normalization & $0.018_{-0.005}^{+0.007}$ & $0.023_{-0.005}^{+0.005}$ & $0.019_{-0.004}^{+0.005}$ \\[1ex]%
\midrule
Statistics & 1756/1597=1.10 & 1794/1602=1.12  &  1790/1601=1.12 \\%
\bottomrule \\
\end{tabular}
\begin{flushleft}
\small
Notes: All models include absorption (\texttt{TBabs}). Normalization is expressed as $10^{-14}\dfrac{\int n_{e}n_{H}dV}{4\pi D^{2}}$, where n$_{e}$ is the electron density of the plasma (cm$^{-3}$), n$_{H}$ is the hydrogen density (cm$^{-3}$), and $D$ (cm) is the distance of the source.
\end{flushleft}
\end{table*}
%


\begin{table*}[]
\centering
\caption{Best fit parameters with $1 \sigma$ errors obtained with a \texttt{TBabs*vapec} model;  O, Ne, Mg, Si, and Fe were left free to vary.
\label{tab:spectra-regions}}
\begin{tabular}{cccc}
\toprule
Region & A & B & C \\
\midrule
$N_{\mathrm{H}}$ (10$^{22}$ cm$^{-2}$) & $0.09_{-0.05}^{+0.05}$ & $0.08_{-0.02}^{+0.02}$ & $0.10_{-0.03}^{+0.02}$  \\[1ex]
kT (keV) &$0.61_{-0.03}^{+0.03}$ &  $0.61_{-0.03}^{+0.03}$  &  $0.60_{-0.02}^{+0.02}$\\[1ex]%
O/O$_{\odot}$ & $1.7_{-0.9}^{+3.2}$ &  $0.9_{-0.3}^{+0.4}$ &  $0.2_{-0.2}^{+0.2}$ \\[1ex]%
Ne/Ne$_{\odot}$ & $0.9_{-0.7}^{+2.4}$ & $0.6_{-0.3}^{+0.4}$ & $0.07_{-0.07}^{+0.2}$ \\[1ex]%
Mg/Mg$_{\odot}$ & $0.05_{-0.05}^{+0.53}$ &  $0.1_{-0.1}^{+0.3}$  & $0.05_{-0.05}^{+0.09}$ \\[1ex]%
Si/Si$_{\odot}$ & $0.7_{-0.7}^{+2.1}$ & $0.7_{-0.4}^{+0.5}$ & $0.6_{-0.3}^{+0.4}$  \\[1ex]%
Fe/Fe$_{\odot}$ & $1.0_{-0.4}^{+1.4}$ & $0.4_{-0.1}^{+0.1}$  &  $0.32_{-0.06}^{+0.09}$ \\[1ex]%
Normalization ($10^{-3}$) &  $0.9_{-0.6}^{+1.3}$ & $4.6_{-1.2}^{+1.5}$ & $8.8_{-2.2}^{+2.4}$\\[1ex]%
\midrule
Statistics & 1770/1602=1.10 & 1749/1602=1.09 &  1733/1602=1.08 \\%
\bottomrule \\
\end{tabular}
\end{table*}

\section{Discussions} \label{sec:summary}
\subsection{SNR classification confirmation}
Detecting a Galactic SNR candidate in a single frequency band may not confirm its nature definitively; data across multiple wavelengths is crucial for the conclusive identification of the candidate as a remnant. Infrared wavelengths help to ascertain whether the radio emission is thermal or nonthermal; X-rays reveal the object's chemical composition and highlight regions where particle acceleration occurs. \g\ shows an extended structure at low X-ray energies surrounded by a radio shell, and it does not present any diffuse emission in the infrared.

The radio flux density has not been determined in the GPM and MGPS-2 data because the signal-to-noise ratio was insufficient for detection. A two-point spectral index was calculated using 1.4 and 2.3\,GHz flux density data, obtaining a value of $\alpha = -0.8 \pm 0.2$. The negative value thus obtained is consistent with nonthermal synchrotron emission, as expected from a shell-type SNR.

\subsection{Distance and age estimates}
Given that \g\ does not have positional coincidences with \textsc{Hi}, \textsc{Hii}, and molecular clouds, providing a distance estimate to the remnant is difficult. However, we can determine a lower and upper limit by applying the studies conducted by \citet{Case1998} on SNRs in the Magellanic Clouds. The authors show that SNRs at 1.4 GHz typically have luminosity values in the range  $5 \times 10^{14} < L_{\rm 1.4\,GHz} < 10^{17}$ W~Hz$^{-1}$. If we suppose that the shell has a luminosity greater than the lower value of this range, using the relation
\begin{equation}
D=\sqrt{\dfrac{L_{\rm 1.4\,GHz}}{4\pi S_{\rm 1.4\,GHz}}},   
\end{equation} 
we can determine its distance from Earth (i.e., $D \gtrsim 700$ pc). Furthermore, if we consider the greatest diameter that a radio SNR could have (i.e., 100\,pc) \citep{Badenes2010}, by applying straightforward calculations we can determine that its distance from Earth must be $D < 3.4$ kpc. Using this result in the formula above, we obtain an upper limit on the luminosity of $L_{\rm 1.4\, GHz} < 1.3 \times 10^{16}$ W~Hz$^{-1}$.

\begin{figure}
    \centering
    \includegraphics[width=1.0\columnwidth]{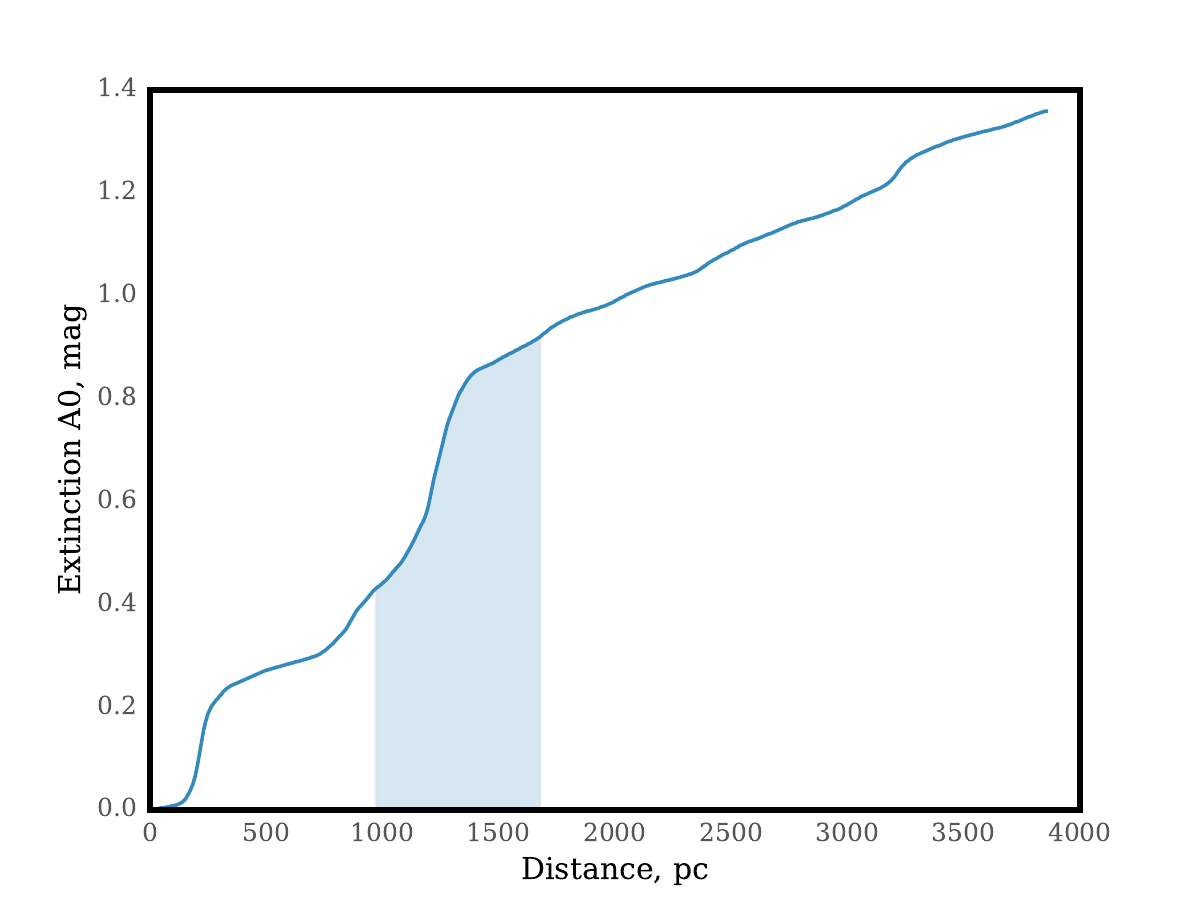}
    \caption{Extinction A0 in the direction of \g\ as provided by the  \textit{Gaia}/2MASS extinction map. The blue shaded area is the distance corresponding to the extinction range derived from hydrogen column density obtained from our X-ray spectral analysis.}
    \label{fig:extinction}
\end{figure}

We also used the results of the X-ray spectral analysis to estimate the distance to the remnant. Although the limited photon statistics do not allow us to constrain the tested spectral models, we estimated the column absorption density $N_{\text{H}}$ toward \g\ to fall within the range of $(0.9 - 1.9) \times 10^{21}$ cm$^{-2}$, considering $1 \sigma$ uncertainties. 
Employing the relation proposed by \cite{Zhu2017},
\begin{equation}
    N_{\text{H}} [\text{cm}^{-2} /A_v ] = (2.08 \pm 0.02) \times 10^{21},
\end{equation}
we derived an anticipated extinction of $A_v \sim (0.4 - 0.9)$. 
This extinction value corresponds to a distance range of (1.0 - 1.7) kpc 
when compared to the \textit{Gaia} data set of \cite{Lallement2019} using the tool available online,\footnote{\url{https://astro.acri-st.fr/gaia_dev/}} as shown in Fig.~\ref{fig:extinction}.

We used the SNR evolutionary model calculator developed by \cite{Leahy2017} to estimate an age limit for \g. The shell is situated close enough to the Galactic plane; consequently, we can assume an ISM density of $1.0 \, \text{cm}^{-3}$ following the studies conducted by \cite{Zhu2017} who analyzed the distribution of the hydrogen in the Galaxy using hundreds of SNRs for which the distance and the column density is well known. Furthermore, we used a standard value for the electron temperature of 100\,K. For a distance of $1.0 < D < 1.7$\,kpc (corresponding to a diameter ranging between 15\,pc and 25\,pc), the interval of possible ages falls between $13.7 < t < 70.1$\;kyr. The transition time between the Sedov-Taylor and radiative phases, in this condition, is estimated to happen at $\sim$13\,kyr, suggesting that the SNR is dominated by radiative mechanisms and that the X-ray map is predominantly composed of the hot gas left behind by the shock wave interaction with the medium. This is further supported by the mixed morphology of \g, which shows a shell structure in the radio and a filled center in X-rays. Additionally, the bright optical filaments discovered by \citet{Fesen2024} also indicate that \g\ is in the radiative--snow-plow phase \citep{Vink2020}.

To have a more accurate estimate of the shell's distance, we also checked whether a pulsar could be associated with it. Due to its proximity to the Galactic plane latitudes, we examined an area within a $\ang{2}$ radius from the remnant's geometric center. This choice aligns with the maximum distance from the remnant's center that a pulsar could have reached of approximately $2\fdg4$ if we consider the edge case of a source at the same distance from Earth as the SNR (1.7\,kpc), with an age of 70.1\,kyr and a maximum proper motion of 1000\,km/s.
We compared it to the Australia Telescope National Facility pulsar catalog v1.70 \citep[][]{Manchester2005}.\footnote{https://www.atnf.csiro.au/research/pulsar/psrcat/} None of the pulsars in the same region have age or proper motion parameters that could be related to a SNR of the size of \g. We performed PARKES follow-up observations in the Ultra-Wideband Low (UWL) search mode using the MEDUSA back-end at 1.5\,GHz to explore the possibility of a pulsar association within the remnant shell without any specific candidate in mind. There is no discernible evidence of a compact object inside the shell at the selected observing frequency. We evaluated the sensitivity of our observations by utilizing the pulsar version of the radiometer equation to determine the limiting flux density \citep[as specified in ][]{Lorimer2004}. The minimum detectable flux density obtained is 0.02\,mJy at a signal-to-noise ratio (S/N) of 5, leading to the exclusion of pulsars at this brightness. From the  \g\ X-ray spectra, we conclude that the SNR progenitor corresponds to a Type Ia supernova, which is in line with the likely absence of a compact remnant associated with it.

\section{Conclusions}\label{sec:conclusions}
The re-discovery of the SNR candidate \g\ in the radio band with the MWA/GPM survey and in X-rays in eROSITA data lets us confirm with a high level of confidence the nature of the object itself. This work successfully demonstrates how a multi-wavelength study of astronomical objects can confirm their classification. Radio observations alone would not have been enough to establish the real origin of the observed diffuse emission. Still, identifying diffuse X-ray emission within the shell and measuring a negative spectral index in archival radio data have provided the required support for a conclusive SNR classification. The independent re-discovery of the remnant in optical by \cite{Fesen2024} strengthens our result.

We will continue to search for more radio and X-ray bright SNR candidates along the plane, identified through combined studies of MWA and eROSITA survey data. Future wide-field low-frequency radio surveys would help reduce the discrepancy between observations and theoretical calculations in the SNR population. The falling spectra of these synchrotron sources make them brighter at low frequencies. This enables us to identify fainter and older remnants that may have a surface brightness that are too low to be detectable at higher radio frequencies. 
The Square Kilometer Array (SKA) is a strong example of instrumentation that would make a difference in supernova remnant detection as it will cover a frequency range of $50 - 350$\,MHz with a resolution of 2\,arcsec and a sensitivity of $\sim 20 \, \mu \text{Jy/beam}$. The shortest baseline of the new interferometer will vary between 45 and 18 meters, respectively, if full stations or a subset are used, leading to a maximum angular scale between 1--3 degrees.

\begin{acknowledgement}
We want to thank the referee for the comments that helped to improve the paper.  
NHW is supported by an Australian Research Council Future Fellowship (project number FT190100231) funded by the Australian Government.
AK thanks Martin G. F. Mayer for the helpful discussions and suggestions.
AK acknowledges support from the International Max-Planck Research School (IMPRS) on Astrophysics at the Ludwig-Maximilians University (IMPRS).
LN acknowledges support from the Deutsche Forschungsgemeinschaft through the grant BE 1649/11-1.

This work has made use of S-band Polarisation All Sky Survey (S-PASS) data and the Continuum HI Parkes All-Sky Survey.
 
This work is based on data from eROSITA, the soft X-ray instrument aboard SRG, a joint Russian-German science mission supported by the Russian Space Agency (Roskosmos), in the interests of the Russian Academy of Sciences represented by its Space Research Institute (IKI), and the Deutsches Zentrum f\"ur Luft- und Raumfahrt (DLR). The SRG spacecraft was built by Lavochkin Association (NPOL) and its subcontractors and is operated by NPOL with support from the Max Planck Institute for Extraterrestrial Physics (MPE). The development and construction of the eROSITA X-ray instrument was led by MPE, with contributions from the Dr. Karl Remeis Observatory Bamberg \& ECAP (FAU Erlangen-Nuernberg), the University of Hamburg Observatory, the Leibniz Institute for Astrophysics Potsdam (AIP), and the Institute for Astronomy and Astrophysics of the University of T\"ubingen, with the support of DLR and the Max Planck Society. The Argelander Institute for Astronomy of the University of Bonn and the Ludwig Maximilians
Universit\"at Munich also participated in the science preparation for eROSITA. The eROSITA data shown here were processed using the eSASS software system developed by the German eROSITA consortium.

This scientific work uses data obtained from Inyarrimanha Ilgari Bundara / the Murchison Radio-astronomy Observatory. We acknowledge the Wajarri Yamaji People as the Traditional Owners and native title holders of the Observatory site. CSIRO’s ASKAP radio telescope is part of the Australia Telescope National Facility (https://ror.org/05qajvd42). Operation of ASKAP is funded by the Australian Government with support from the National Collaborative Research Infrastructure Strategy. ASKAP uses the resources of the Pawsey Supercomputing Research Centre. Establishment of ASKAP, Inyarrimanha Ilgari Bundara, the CSIRO Murchison Radio-astronomy Observatory and the Pawsey Supercomputing Research Centre are initiatives of the Australian Government, with support from the Government of Western Australia and the Science and Industry Endowment Fund. This paper includes archived data obtained through the CSIRO ASKAP Science Data Archive, CASDA (https://data.csiro.au).
\end{acknowledgement}

\bibliographystyle{aa}
\bibliography{Biblio}

\begin{appendix}
\section{Polarization}
The radiation coming from SNRs is linearly polarized. Ideally, it could be possible to determine the orientation of the magnetic field in the plane of the sky, as has been done by \cite{Shanahan2022}. However, Faraday rotation changes the electric field's orientation due to the interaction with the ISM present in the same sky area and along the line of sight, changing the orientation of the signal we receive. In addition, Faraday rotation is inversely proportional to the frequency, becoming important at low radio frequencies. 

\begin{figure}
\centering
\includegraphics[width=1.0\linewidth]{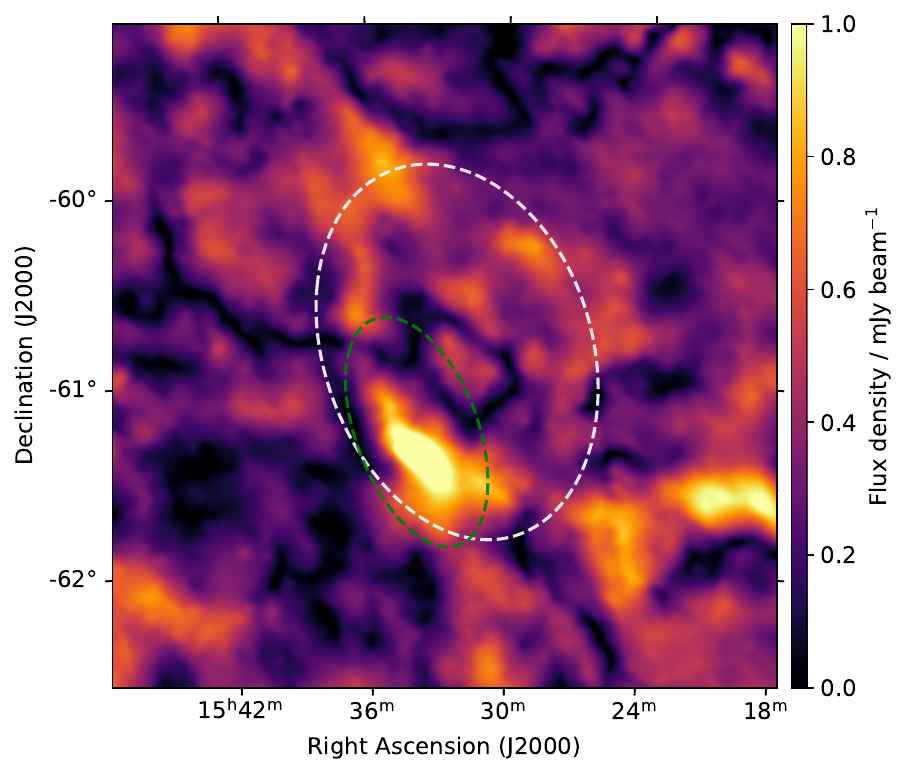}
\caption{Total linear polarization of $3\fdg5 \times 3\fdg5$ deg of the region surrounding \g\, estimated from Stokes Q and U images of S-PASS. The white dashed line highlights the remnant shell, and the green dashed line corresponds to the polarized part of the shell we examined.}
\label{fig:Polarization}
\end{figure}

We examined by eye the remnant's Stokes Q and U images to verify if linear polarization, respectively, is present at $\pm \ang{90}$ and $\pm \ang{45}$ angles. We could not determine the magnetic field vectors and strength because we only had data at a single frequency. We could, instead, morphologically identify a region of linear polarization in coincidence with the bottom-left branch of \g\ shell. We determine the total polarization fraction (see \fig~\ref{fig:Polarization}) as: $p = \dfrac{\sqrt{U^2 + Q^2}}{I}$, obtaining a value of $p = 15\%$.

\section{RACS}\label{racs}
\begin{figure*} 
\centering
\includegraphics[width=1.0\linewidth]{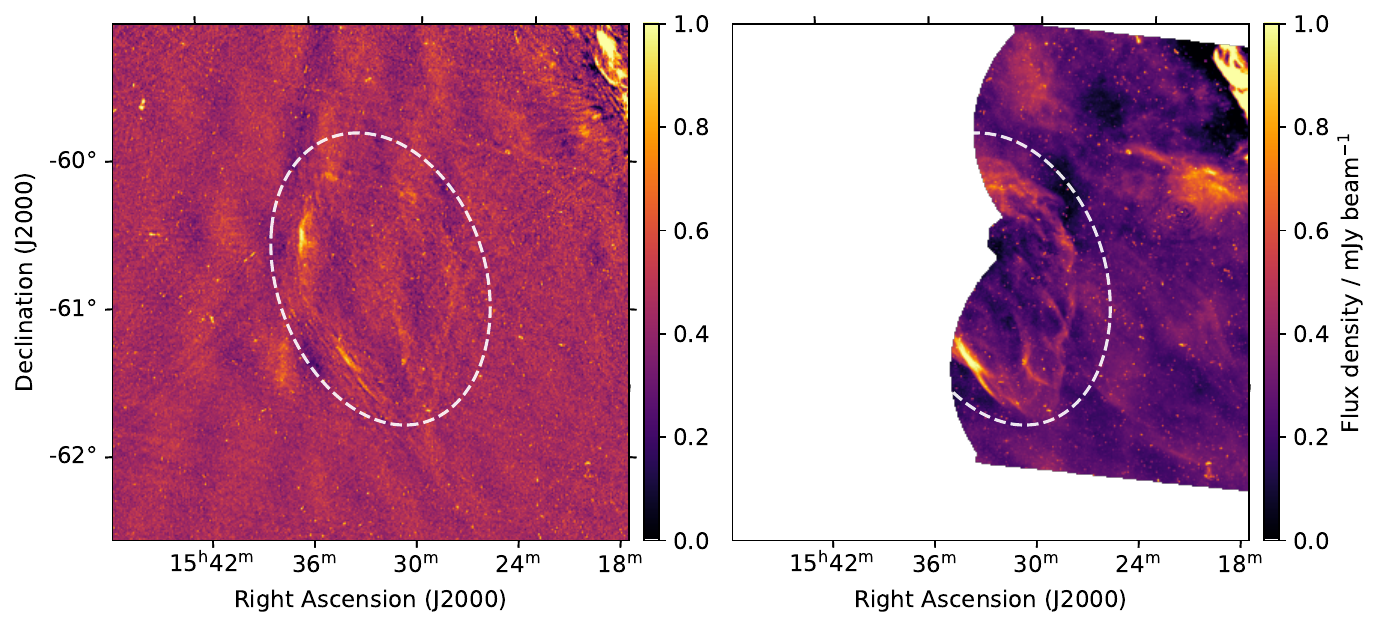}
\caption{$3\fdg5 \times 3\fdg5$ deg of the region surrounding \g\ as seen at $\sim 888$\,MHz by RACS in the left panel and at 944\,MHz by EMU in the right panel.}
\label{fig:RACS} 
\end{figure*}

\g's shell is also visible in the first data release of RACS at a frequency of $\sim 888$\,MHz. The survey covers more than $34000$\,degrees$^2$ with $- \ang{90} < \delta < + \ang{41}$; its resolution and sensitivity, respectively, take value of $\sim 15\arcsec$ and $0.25\,\text{mJy beam}^{-1}$. The mean total intensity map has been downloaded\footnote{https://data.csiro.au/collection/csiro:46533} to create a $3\fdg5 \times 3\fdg5$ cutout around the SNR candidate. The structure can be seen in the left panel of \fig~\ref{fig:RACS}.

\section{EMU}\label{emu}
EMU is a wide-field radio continuum survey collected using the new Australian Square Kilometre Array Pathfinder (ASKAP) telescope. It is currently observing the southern sky at 944\,MHz south of declination $\ang{+30}$. It observes the sky with a resolution of $15\arcsec$ and a noise of $25 \, \mu\text{Jy/beam}$. It is the deepest survey we have so far, making it possible to distinguish the fine details of the internal structure of \g\ as shown in the right panel of \fig~\ref{fig:RACS}. It is an ongoing project, so the sky region where \g\ resides has not been completely observed yet, but the first image made available is already promising. 

\section{XMM-Newton observation}
\label{sec.xmm}
Within a circle of $90\arcmin$ diameter covering the SNR, $\sim 180$ X-ray point sources are detected in the eRASS:4 data. We inspected the \textit{XMM-Newton} data archive for observations covering the SNR area. On March $21^{\rm st}$, 2012 a $\simeq 11.6$\;ks observation centered at RA = 15:33:20.710, Dec = $-60$:14:11.30 covered a $42\arcmin\times42\arcmin$ area in the northern part of the remnant. A total of 8 point sources are reported in the 4XMM-DR13 catalog \citep{Webb+2020}. A few of them are also visible in the eROSITA images. The two strongest sources in the field are 4XMM J153254.1$-$601418 (S1) and 4XMM J153323.4$-$601019 (S2), with a cataloged 0.2--10 keV flux of $(1.24\pm 0.18)\times 10^{-13}$ erg\;cm$^{-2}$\;s$^{-1}$, and $(1.38\pm 0.34)\times 10^{-13}$ erg\;cm$^{-2}$\;s$^{-1}$, respectively.

\cite{Tranin+2022} performed a Bayes classification of the \textit{XMM-Newton} sources, and the produced catalog identifies both these sources as stars at a confidence level $>99\%$\footnote{https://vizier.cds.unistra.fr/viz-bin/VizieR-3?-source=J/A A/657/A138/table7}. Considering the $\sim 2\arcsec$ positional uncertainty of the \textit{XMM-Newton} detection, we identify them in the \textit{Gaia} DR3 catalog \citep{gdr3cat2022}: S1 is a $13.1$ Gmag M3-type star at a distance of $\sim 67$ pc and a high proper motion of $\sim 118$ mas\;yr$^{-1}$, whereas S2 is a $10.8$ Gmag rK3III-type giant star at a distance of $\sim 1$ kpc.
High background level affecting the whole observation prevents us from using these data for diffuse emission studies.

\end{appendix}

\end{document}